\documentclass[manuscript]{aastex}

\shorttitle{{\it In situ} lightcurves}
\shortauthors{Short}

\begin{document}

%% LaTeX will automatically break titles if they run longer than
%% one line. However, you may use \\ to force a line break if
%% you desire.

\title{{\it In situ} exo-planet transit lightcurve modelling with the Chroma+ suite}

%% Use \author, \affil, and the \and command to format
%% author and affiliation information.
%% Note that \email has replaced the old \authoremail command
%% from AASTeX v4.0. You can use \email to mark an email address
%% anywhere in the paper, not just in the front matter.
%% As in the title, use \\ to force line breaks.

\author{C. Ian Short}
\affil{Department of Astronomy \& Physics and Institute for Computational Astrophysics, Saint Mary's University,
    Halifax, NS, Canada, B3H 3C3}
\email{ian.short@smu.ca}

%% Notice that each of these authors has alternate affiliations, which
%% are identified by the \altaffilmark after each name.  Specify alternate
%% affiliation information with \altaffiltext, with one command per each
%% affiliation.

%\altaffiltext{1}{Visiting Astronomer, Cerro Tololo Inter-American Observatory.
%CTIO is operated by AURA, Inc.\ under contract to the National Science
%Foundation.}
%\altaffiltext{2}{Society of Fellows, Harvard University.}
%\altaffiltext{3}{present address: Center for Astrophysics,
%    60 Garden Street, Cambridge, MA 02138}
%\altaffiltext{4}{Visiting Programmer, Space Telescope Science Institute}
%\altaffiltext{5}{Patron, Alonso's Bar and Grill}

%% Mark off your abstract in the ``abstract'' environment. In the manuscript
%% style, abstract will output a Received/Accepted line after the
%% title and affiliation information. No date will appear since the author
%% does not have this information. The dates will be filled in by the
%% editorial office after submission.

\begin{abstract}

We have added to the Chroma+ suite of stellar atmosphere and spectrum modelling 
codes the ability to synthesize the exo-planet transit lightcurve for
planets of arbitrary size up to 10\% of the host stellar radius, and arbitrary
planetary and stellar mass and orbital radius (thus determining orbital velocity) 
and arbitrary orbital inclination.  The lightcurves are computed {\it in situ},
integrated with the radiative transfer solution for the radiation field emerging
from the stellar surface, and there is no limb-darkening parameterization.
The lightcurves are computed for the 
Johnson-Bessel photometric system $UBVRIHJK$.  We describe our method of
computing the transit path, and the reduction in flux caused by occultation,
and compare our lightcurve to an analytic solution with a four-parameter 
limb-darkening parameterization for the case of an edge-on transit of the Sun by Earth.
This capability has been added to all ports and variations, including the
Python port, ChromaStarPy, and the version that interpolates among
the fully line-blanketed ATLAS9 surface intensity distributions, ChromaStarAtlas.
All codes may be accessed at www.ap.smu.ca/OpenStars and at GitHub
(github.com/sevenian3). 
   
\end{abstract}

%% Keywords should appear after the \end{abstract} command. The uncommented
%% example has been keyed in ApJ style. See the instructions to authors
%% for the journal to which you are submitting your paper to determine
%% what keyword punctuation is appropriate.


\section{Introduction}

  Transit lightcurve analysis has become an important tool in determining the properties of exo-planets
and of their orbital parameters, and of the host star.  They pose an interesting inverse-problem and much effort has gone into
extracting information about the system from the detected lightcurve, and most of these methods rely in one way or
another on a parameterization of the host star's limb-darkening profile.  The limb-darkening coefficients (LDCs) that 
parameterize a limb-darkening law are wavelength- and band-pass- sensitive and must be determined for each
photometric system, and for each set of host star parameters.  Moreover, limb-darkening laws are necessarily an approximation to the real
variation of specific intensity with angle of emergence from the host star's surface, $I_\lambda(\cos\theta)$.

\paragraph{}

We have implemented a complementary forward-modelling approach by incorporating the calculation of transit lightcurves, $F_{\rm band}(t)$, in the Chroma+ suite of stellar atmosphere and spectrum modelling codes for the
Johnson-Bessel
$U_{\rm x}BVRI$ \citep{johnson66} and Johnson $HJK$ \citep{HJK} photometric systems.  
Our procedure computes $F_{\rm band}(t)$ {\it in situ} 
because it is integrated with the radiative transfer solution for the emergent surface intensity 
$I_\lambda(\tau=0, \cos\theta)$ for the atmospheric structure, where $\tau$ is any vertical optical depth scale 
increasing inward.  Therefore, there is no limb-darkening
parameterization.  This approach was also taken by \citet{neilson} to evaluate the accuracy of LDC-based
lightcurve analysis using $I_\lambda(\tau=0, \cos\theta)$ distributions computed with the plane-parallel
and spherical versions of the ATLAS9 \citep{castellik06} stellar atmosphere and spectrum modelling suite.
Our simulated $F_{\rm band}(t)$ curves may be used to evaluate the accuracy of 
LDC-based inverse methods, as well as for the forward modelling of observed $F_{\rm band}(t)$ signals
with a grid that explores host-star, planetary, orbital parameter, and orientation space.

\paragraph{}

We have added $F_{\rm band}(t)$ calculation to the entire Chroma+ suite, including the Python implementation
ChromaStarPy (CSPy, \citet{shortbb18}), which provides for fast modelling and analysis in a Python IDE, and ChromaStarAtlas
\citep{shortb18}, in which the $I_\lambda(\tau=0, \cos\theta)$ distribution being occulted is the fully line blanketed
distribution interpolated with the public ATLAS9 $I_\lambda(\tau=0, \cos\theta)$ distributions of 
\citet{castellik06}.

\section{Method}

The radiative transfer procedure of the Chroma+ suite computes the emergent monochromatic surface specific
intensity distribution, $I_\lambda(\tau=0, \cos\theta)$, for a set of direction angles, $\{\cos\theta\}$,
with respect to the local stellar surface normal that have a Gauss-Legendre distribution in the 
$\cos\theta$ range $[1, 0]$,
over a wide range of $\lambda$ from the UV to the IR with equal $\log\lambda$ spacing supplemented with
{\it ad hoc} additional $\lambda$ points for spectral lines.  This is the
$I_\lambda(\tau=0, \cos\theta)$ distribution that is occulted as an exo-planet transits the host star as
seen by an observer on Earth.

\subsection{Assumptions}

We adopt the following simplifying assumptions for the planetary system:  
1) The exo-planet orbital radius, $R_{\rm orb}$, is large enough compared to
the stellar radius, $R$, that the
transit path is a chord in the plane of the sky, 2) The exo-planet's orbit is Copernican so that, along with 
assumption 1), the component of orbital 
velocity, $v_{\rm orb}$, in the plane-of-the-sky is constant during transit and is equal to $v_{\rm orb}$, 
3) The planet's radius, $r$, is small enough to occult only one 
$\Delta\cos\theta$ substellar-centric 
annulus in the discretization in the plane of the sky of the host stellar atmosphere radiation field at any 
time $t$, 
4) Only transits in which the entire projected area of the planetary disk is occulting at mid-transit are of interest,  
5) The planet has a specific intensity of zero, 6) The distance to the system, $d$, is large compared to $R_{\rm orb}$ so that
the occulted flux may be calculated at the stellar surface, and so that, along with assumptions 1) and 2), the 
transit velocity is equal to $v_{\rm orb}$.  Assumption 3) is the ''small planet approximation'' investigated by
\citet{mandel02} and corresponds to $r/R \lesssim 0.1$.  Assumption 4) is consistent
with Assumption 3), and disregards grazing transits, which are less detectable.

\subsubsection{Inputs}

In addition to the host stellar parameters required for static $1D$ horizontally homogenous plane-parallel
modelling of the host stellar atmosphere ($T_{\rm eff}, \log g, [{{\rm A} \over {\rm H}}], \xi_{\rm T}$),
the procedure also requires the radius of the exo-planet orbit, $R_{\rm orb}$, the radius of the exo-planet, $r$, and the inclination
of the planetary orbital axis with respect to the line-of-sight, $i$.  As part of the established Chroma+ 
modelling procedure, the user also specifies an input stellar mass, $M$, which the Chroma+ codes combine with 
the input $\log g$ value to compute the host star's radius, $R$.  We assume that the system is Keplerian
($m_{\rm planet} << M$) so that $v_{\rm orb}$ is found from $v^2_{\rm orb} = GM/R_{\rm orb}^2$.      

\subsection{The transit path}

Let $S$ be the substellar point,
$P$ be the position of the planet's centre at any time $t$ during transit, and $P_0$ be the location of $P$ at mid-transit,
all projected into the plane of the sky,
so that a line extending from $S$ through
$P_0$ bisects the transit path chord, and let $t$ be the time coordinate with $t=0$ at mid-transit when $P = P_0$.
We relate the transit path $P(t)$ to the spherical polar coordinate $\theta$ in the standard discretization 
of the stellar atmospheric radiation field geometry, in which the positive polar axis $z$ extends from the centre 
of the star through the point $S$ to the observer, with the following procedure.  
We first compute the impact parameter, $b_{\rm min}$, which is the length of the segment $SP_0$, 
corresponding to mid-transit, as $b_{\rm min} = R_{\rm orb}\sin (\pi-i)$.  Assumption 4) corresponds to
the condition that $b_{\rm min} < R - r$.  The corresponding 
minimum value of $\theta$ along the transit path is then found from
 $\sin\theta_{\rm min} = b_{\rm min}/R$.  
For each {\it a priori} $\theta$ value in the discretization of the stellar radiation field, 
the separation of $P$ and $S$, $b(\theta)$, is 
found from $b = R\sin\theta$.  Then, defining $\Delta x$ to be the length of the segment $P_0P$, the linear distance traversed by the planet 
at time $t$, $\Delta x$ is found from $\Delta x = \sqrt{b^2-b^2_{\rm min}}$, the value of $t(\theta)$
is $\Delta x/v_{\rm orb}$, and the set $\{\theta_{\rm i}(t_{\rm i})\}$ determines which $I_\lambda(\cos\theta)$
beams are occulted as a function of time.  These $\theta(t)$ values are for a half-transit, and the other half 
of the transit path is found by reflection about $P_0$ under the assumption that the stellar radiation field is axi-symmetric
about $z$.  

\paragraph{}

Our $\{\cos\theta_{\rm i}\}$ set is that of a Gauss-Legendre quadrature on the interval $[-1, 1]$, consistent with standard practice in stellar
radiation field modelling.  The advantage here is that the points $P$ are distributed so that the transit lightcurve is 
sampled with increasing density as the light varies more rapidly with $x$ along the transit path as the transiting 
planet approaches the stellar limb and egress.   

\subsection{Occulted flux}

\subsubsection{Interior of lightcurve}

Under the assumption that $d >> R$ so that the monochromatic flux at Earth, $f_\lambda$, only consists of parallel
beams emerging from projected annuli at the stellar surface, the un-occulted flux at the stellar surface
($d=R$) is approximated with our $\{\cos\theta_i\}$ grid and out-going $I_\lambda(\tau=0, \cos\theta_i)$ 
beams in the $\cos\theta$ range $[0, 1]$ as

\begin{equation}
F_\lambda = \sum^{N/2+1}_{i=0} w_i I_\lambda\cos\theta_i 
\end{equation} 

 where $\{w_i\}$ is the set of Gauss-Legendre quadrature weights for the zero-positive subset of an $N$-point quadrature of 
odd $N$ in the range
$[0, 1]$.  The solid angle subtended by the planet for an observer at $d=R$ is ${\rm d}\omega = \pi(r/R)^2$.
Therefore, the flux occulted by the planet at the stellar surface, $\Delta F_{\lambda, i}$, when $P$ is at 
polar angle $\theta_i$ on the transit path, may be calculated as 

\begin{equation}
\label{eqn1}
\Delta F_{\lambda, i} = {\rm d}\omega I_\lambda(\cos\theta_i)  
\end{equation}

Because $\Delta F_{\lambda, i} << F_\lambda$, the occulted stellar flux during transit, 
$F^{\rm T}_{\lambda, i} = F_\lambda - \Delta F_{\lambda, i}$, 
is calculated for each $\theta_i$ on the transit path as 

\begin{equation}
\label{eqn2}
\log F^{\rm T}_{\lambda, i} = \log F_\lambda + \log\left(1 - \exp(\log\Delta F_{\lambda, i} - \log F_\lambda) \right) 
\end{equation}

and all $F$ values are represented as double precision floating point data-type. 

\subsubsection{Ingress and egress}

The $F^{\rm T}_{\lambda, j}$ variation during egress is modelled with a three-point approximation $\{P_j\}$, $j = 1$ to $3$, corresponding
to positions $P$ on the transit path of $b_j$ equal to $R-r$, $R$, and $R+r$ that span the stellar limb.  
For each $P_j$ position, the corresponding $\Delta x_j$ value is found from
$\Delta x_j = \sqrt{(b^2_j - b^2_{\rm min})}$ and then $t_j$ values from $\Delta x_j/v_{\rm orb}$.  
For $P_1$, $\Delta F_{\lambda, j=1}$ is found from Eq. \ref{eqn1} with $i=N$, the
smallest value in the $\{\cos\theta_i\}$ quadrature set corresponding to the annulus nearest the stellar limb.  
For $P_3$, $\Delta F_{\lambda, j=3} = 0$.  
For $P_2$, close to mid-egress, we approximate ${\rm d}\omega$ as the solid angle subtended by 
a sector of the planet's projected circular area overlapping the stellar disk equal to 
$(2\phi/2\pi)\pi r^2 = \phi r^2$,
with $\phi$ found from $\tan\phi = R/r$, and then compute $\Delta F_{\lambda, j=2}$ and $F^{\rm T}_{\lambda, j=2}$ from Eqs. 
\ref{eqn1} and \ref{eqn2}.  
The $F^{\rm T}_{\lambda, j}$ values during ingress are
then found by reflection about $P_0$.    

\section{Results and Discussion}

 The $ F^{\rm T}_{\lambda, i}$ values are used to compute the relative change in band-integrated flux, 
$F^{\rm T}_{{\rm band}, i}/F_{\rm band}$,
for the $\{t_i\}$ values using the synthetic photometry module of the Chroma+ suite \citep{shortbb18}
for the Johnson-Bessel $UBVRIHJK$ bands. 
 In Fig. \ref{f1} we show the $F^{\rm T}_{{\rm band}, i}/F_{\rm band}$
{\it vs.} $t$ curves for the $UBVRIK$ bands for a CSPy model of the Sun 
($T_{\rm eff}/\log g/[{{\rm A}\over{\rm H}}]/\xi_{\rm T} = 5777/4.44/0.0/1.0$) 
being transited by a planet of Earth's $r$ and $R_{\rm orb}$ values with
$i = \pi$ RAD (edge-on).

\subsection{Comparison to limb-darkened lightcurves}

\begin{figure}
\plotone{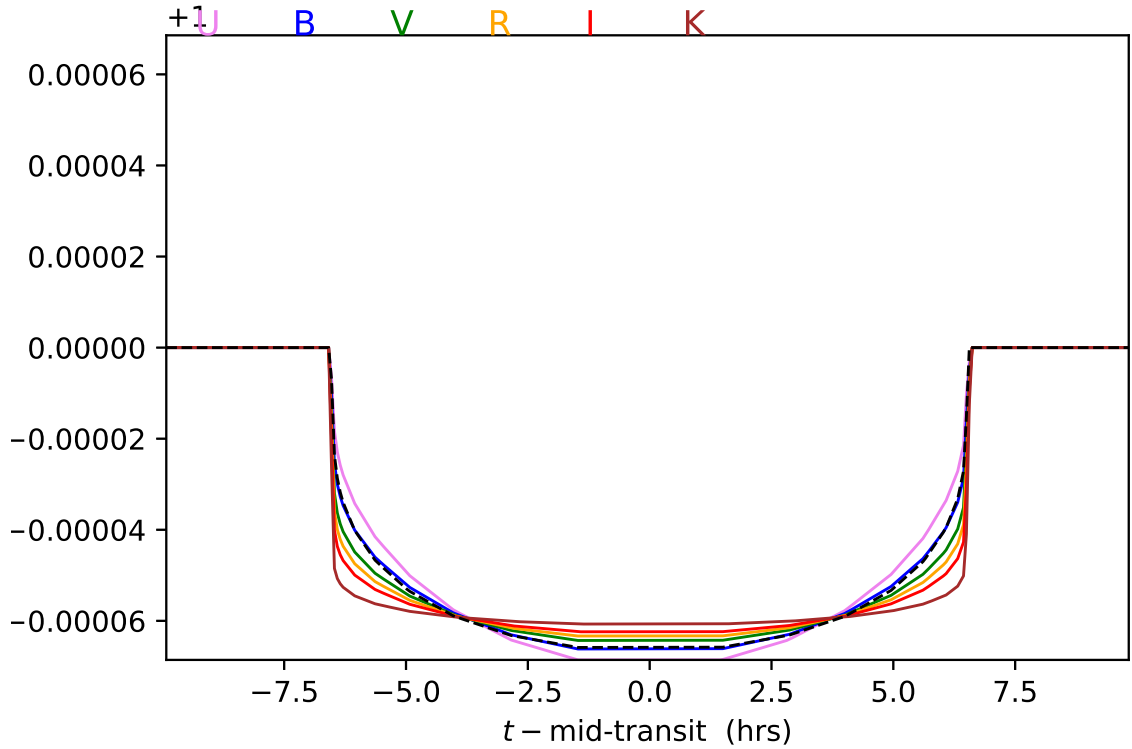}
\caption{$F^{\rm T}_{{\rm band}, i}/F_{\rm band}$ {\it vs.} $t$ curves for the 
Johnson-Bessel $UBVRIK$ bands, calculated at 28 points 
(solid lines) for the case of a solar
host star and a planet of Earth's radius.  A comparable analytic $V$-band interior
lightcurve based
on the four-parameter limb-darkening law of \citet{claret00} is also
included for the adjustment parameter $f=2/\pi$ (dashed line, see text). 
  \label{f1}
}
\end{figure}

We calculate analytically an independent $V$-band lightcurve interior, neglecting ingress and egress,
 for an edge-on transit of a solar-like model from the ATLAS9 atmospheric model 
grid of ($T_{\rm eff}/\log g/[{{\rm A}\over{\rm H}}]/\xi_{\rm T} = 5750/4.5/0.0/1.0$), based on the 
four-parameter second order limb-darkening parameterization of \citet{claret00}, 
$I_{V, {\rm LDC, i}}(\cos\theta_{\rm i}) = 1 - \sum_{n=1}^4 (1-a_n\cos^{n/2}\theta_{\rm i})$ with 
$\{a_n\} = \{0.5169, -0.0211, 0.6944, -0.3892\}$.  We calculate the analytic lightcurve for a planet
of Earth's $R_{\rm orb}$ value using a slightly
modified form of the formula
of \citet{mandel02} for their case of a ''small planet'' ($r/R \lesssim 0.1$) and
the entire projected planetary disk occulting the star, 

\begin{equation}
F^{\rm T}_{V, {\rm LDC, i}}/F_{V, {\rm LDC}} = 1 - fr^2I_{V, {\rm LDC, i}}(\cos\theta_{\rm i})/4R^2\sum_{n=0}^4 a_n/(n+4) 
\end{equation}

where $a_0$ may be found from $1-\sum_{n=1}^4 a_n$.  The modification is the factor $f$, which
allows us to adjust this analytically calculated reduction in relative flux during transit. 
In Fig. \ref{f1} we also show the $F^{\rm T}_{V, {\rm LDC, i}}/F_{V, {\rm LDC}}$ curve for the case of $f = 2/\pi$.

\section{Implementation in CSPy}

Transit set-up is controlled with the addition of the ''rOrbit'' and ''rPlanet'' settings in the Input.py command file
to set the values of $R_{\rm orb}$ and $r$, respectively.
Additionally, the transit is controlled by a number of previously established settings that have other
purposes: the ''logg'' and ''massStar'' settings are used to compute $R$ and $V_{\rm orb}$, and the ''rotI'' setting
for rotational broadening is used to compute $b_{\rm min}$. 

\paragraph{} 
  
    Currently, CSPy's radiation field discretization uses the 11 zero-positive abscissae of a 21-point
Gauss-Legendre quadrature to sample the $\theta$ polar angle coordinate, and thus the $b$ offset 
from the substellar point, and that is the maximum number of points sampling a half-transit 
for the case of $i=\pi/2$ RAD ($b_{\rm min}=0$).  The full interior light curve is sampled with twice this number of points (22), and the three-point treatment of ingress and egress bring the total number of
points sampling the entire light curve to 28, including the two bracketing un-occulted points. 
This relatively modest number has been chosen because responsiveness in a Python IDE is a priority that
distinguishes CSPy from more realistic FORTRAN atmospheric and spectrum modelling codes.
The Thetas.thetas() module in the Chroma+ suite is set up so that it is straightforward to change the
order of the Gauss-Legendre quadrature and, thus, the number of points sampling the lightcurve.    

%% If you wish to include an acknowledgments section in your paper,
%% separate it off from the body of the text using the \acknowledgments
%% command.

%% Included in this acknowledgments section are examples of the
%% AASTeX hypertext markup commands. Use \url without the optional [HREF]
%% argument when you want to print the url directly in the text. Otherwise,
%% use either \url or \anchor, with the HREF as the first argument and the
%% text to be printed in the second.

\acknowledgments

\clearpage

%% Tables may also be prepared as separate files. See the accompanying
%% sample file table.tex for an example of an external table file.
%% To include an external file in your main document, use the \input
%% command. Uncomment the line below to include table.tex in this
%% sample file. (Note that you will need to comment out the \documentclass,
%% \begin{document}, and \end{document} commands from table.tex if you want
%% to include it in this document.)

%% \input{table}

%% The following command ends your manuscript. LaTeX will ignore any text
%% that appears after it.


\begin{thebibliography}{}

%\bibitem[Bautista {\it et al.} (1998)]{bautista}Bautista, M. A., Romano, P. \& Pradhan, A. K., 1998, \apjs, 118, 259
%\bibitem[Box \& Muller (1958)]{boxmuller}Box, G.E.P. \& Muller, M.E., 1958, Annals Math. Stat., 29, 610
\bibitem[Castelli \& Kurucz (2006)]{castellik06} Castelli \& F. Kurucz, R. L., 2006, \aap, 454, 333
%\bibitem[Castelli \& Kurucz (2004)]{castellik04} Castelli \& F. Kurucz, R. L., 2004, Proceedings of the IAU Symp. No 210, Modelling of Stellar Atmospheres, eds. N.E. Piskunov, W.W. Weiss, and D.F. Gray, poster A20
\bibitem[Claret (2000)]{claret00}Claret, A., 2000, \aap, 363, 1081
\bibitem[Johnson (1965)]{HJK} Johnson, H., L., 1965, \apj, 141, 923
\bibitem[Johnson {\it et al.} (1966)]{johnson66} Johnson, H. L., Mitchell, R. I., Iriarte, B. \& Wisniewski, W. Z., 1966, Comm. Lunar Planet. Lab., 4, 99
%\bibitem[Cox (2002)]{allens}Cox, A.N., Ed., 2002, {\it Allen's Astrophysical Quantities}, Fourth Ed., Springer 
%\bibitem[Gray (2005)]{gray}Gray, D.F., 2005, {\it The Observation and Analysis of Stellar Photospheres}, Third Ed., Cambridge University Press
%\bibitem[Gray (1988)]{grayfgk}Gray, D.F., 1988, {\it Lectures on spectral-line analysis : F, G, and K stars}, Arva, Ont. : The Publisher 
%\bibitem[Hauschildt {\it et al.} (1999)]{phoenix}Hauschildt, P.H., Allard, F., Ferguson, J., Baron, E. \& Alexander, D.R., 1999, \apj, 525, 871
%\bibitem[Kramida {\it et al.} (2015)]{nist}Kramida, A., Ralchenko, Yu., Reader, J., and NIST ASD Team, 2015, NIST Atomic Spectra Database (ver. 5.3), [Online]. Available: http://physics.nist.gov/asd [2015, November 26]. National Institute of Standards and Technology, Gaithersburg, MD.
%\bibitem[Kurucz (2014)]{kurucz14} Kurucz, R.L., 2014, Determination of Atmospheric Parameters of B-, A-, F- and G-Type Stars. Series: GeoPlanet: Earth and Planetary Sciences, Eds. E. Niemczura, B. Smalley and W. Pych, Springer International Publishing (Cham), p. 25
%\bibitem[Kurucz (1992)]{atlas9}Kurucz, R.L., 1992, Rev. Mex. Astron. Astrofis., 23, 181
%\bibitem[Lindstrom \& Mallard (2017)]{antoine}Linstrom, P.J. \& Mallard, G. Eds., NIST Chemistry WebBook, NIST Standard Reference Database Number 69, National Institute of Standards and Technology, Gaithersburg MD, 20899, doi:10.18434/T4D303, (retrieved April 12, 2017)
\bibitem[Mandel \& Agol (2002)]{mandel02}Mandel, K. \& Agol, E., 2002, \apjlett, 580, L171
%\bibitem[Mishenina {\it et al.} (2016)]{rgbheavy}Mishenina, T., Kovtyukh, V., Soubiran, C., Adibekyan, V. Zh., 2016, \mnras, 462, 1563
\bibitem[Neilson {\it et al.} (2017)]{neilson}Neilson, H.R., McNeil, J.T., Ignace, R. \& Lester, J.B., 2017, \apj 845, 65
%\bibitem[Newton {\it et al.} (2015)]{rgbir}Newton, E. R., Charbonneau, D., Irwin, J., Mann, A. W., 2015, \apj, 800, 85
\bibitem[Short \& Bayer (2018)]{shortb18} Short, C.I. \& Bayer, J.H.T., 2018, arXiv:1805.03674
\bibitem[Short, Bayer \& Burns (2018)]{shortbb18} Short, C.I., Bayer, J.H.T. \& Burns, L.M., 2018, \apj, 854, 82
%\bibitem[Short (2016)]{gss16}Short, C.I., 2016, \pasp, 128, 104503 (S16)
%\bibitem[Short (2014)]{methods}Short, C.I., 2014b, arXiv:1409.1893  
%\bibitem[Short {\it et al.} (1999)]{shortPHX}Short, C.I., Hauschildt, P.H. \& Baron, E., 1999, \apj, 525, 375
%\bibitem[Sneden {\it et al.} (2012)]{moog}Sneden, C., Bean, J., Ivans, I., Lucatello, S. \& Sobeck, J., 2012, Astrophysics Source Code Library, ascl:1202.009
%\bibitem[Zeidler-K.T. \& Koester (1982)]{jola}Zeidler-K.T, E.M. \& Koester, D., 1982, \aap, 113, 173
\end{thebibliography}
\end{document}